\documentclass[aps,prc,preprint,superscriptaddress,showpacs,floatfix]{revtex4}

\usepackage{CJK}
\usepackage{graphicx}

\begin{document}

\title{Shape coexistence close to $N=50$ in the neutron-rich isotope $^{80}$Ge investigated by IBM-2}

\begin{CJK*}{GBK}{song}

\author{ZHANG DaLi}
\email[Corresponding author:] {zdl@zjhu.edu.cn}
\affiliation{Department of physics, Huzhou University,
Huzhou 313000, Zhejiang, China}
\author{MU ChengFu}
\affiliation{Department of physics, Huzhou University,
Huzhou 313000, Zhejiang, China}

\date{\today}
\begin{abstract}
The properties of the low-lying states, especially the relevant shape coexistence in $^{80}$Ge close
to one of most neutron-rich doubly magic nuclei at $N=50$ and $Z=28$ have been investigated within the
framework of the proton-neutron interacting model (IBM-2). Based on the fact that the relative energy of the $d$
neutron boson is different from proton bosons', the calculated energy levels of low-lying states, $E2$
transition strengths can reproduce the experimental data very well. Particularly, the first excited
state 0$^+_2$ is reproduced quite nicely, which is intimately related with the shape coexistence phenomenon. And the
$\rho^2(E0, 0^+_2\rightarrow0^+_1)$ transition strength has been predicted. The experimental data and theoretical
results indicate that both collective spherical and $\gamma$-soft vibration structures coexist in $^{80}$Ge.
\end{abstract}

\maketitle

\end{CJK*}



\newpage

\section{Introduction}
\label{s1}

Shape coexistence is a peculiar nuclear phenomenon when two or more states occur
in the same nuclei within a very narrow energy range at low excitation energy \cite{Poves01}.
Shape coexistence phenomenon is often found close to or at the shell closures, where
deformed intruder configurations coexist with spherical shapes based on multiparticle-hole
excitations across the closed shell in the nuclear chart, from light nuclei to heavy
nuclei \cite{Heyde02,Gade03,Nowacki04}. The presence of low-lying 0$^+$ states as
the first excited state in even-even nuclei is one of the signatures of shape
coexistence \cite{Heyde02,Andreyev05}, which plays an important role in our understanding
for the shape changes of the nuclear structure in the exotic nuclei.\

In recent years, the advent of radioactive isotopes beams have been developed,
which gives the access to exotic nuclei far from stability in both the neutron-deficient
and neutron-rich regimes \cite{Poves01,Gorgen06,Liu07,Dong08,Bai15,Sun16}. In the neutron-rich
nuclei, the empirical evidence of shape coexistence has been observed
along $N=20$, $N=28$ and the subshell gap $N=40$, see Refs.[2,3] for a review. A lot of
theoretical works have been developed to investigate shape coexistence and
shape phase transition, for example, the interacting-boson model \cite{Iachello1998,Liu2006},
the shell model \cite{Hasegawa2007} and the projected shell model \cite{Sun2009,Liu2011},
the self-consistent relativistic mean-field theory \cite{Ren2002}.
Nowadays shape coexistence in nuclei close to the supposedly doubly magic nucleus
$^{78}_{28}$Ni$_{50}$ is the focus of intense experimental and theoretical research
(cf., for example \cite{Nowacki04,Hagen09,Gottardo11} and references therein),
because the study of shape coexistence in this region will help us to differentiate the
single-particle effect from the quadrupole collective motion across $N=50$.
More recently the technique of $\beta$-delayed electron-conversion
spectroscopy is applied to study $^{80}$Ge nucleus. In Ref.~\cite{Gottardo11},
an electric monopole $E0$ transition is observed for the first time, which points
to an intruder 0$^+_2$ state at 639(1) keV. The new state 0$^+_2$ is much lower than
the 2$^+_1$ level in $^{80}$Ge, this characteristic implies there might exist the
effect of shape coexistence near the most neutron-rich doubly magic
nucleus at $N=50$ and $Z=28$, giving an insight into the mechanism of
shape coexistence close to the neutron major shell closure at $N=50$. \

It is well known that the low-lying structure of Ge isotopes display the trends
of coexistence of different shapes along the long isotopic chain, characterized
by prolate-oblate and spherical-deformed competition. Close to the $\beta$-stability
line, the shape transition of Ge isotopes is a drastic evolution from
nearly spherical in $^{72}$Ge to slight prolate in $^{74}$Ge or even triaxiality in
$^{76,78}$Ge \cite{Padilla-Rodal13,Sheng14,Zhang15,Zhang20172} and $^{84,86,88}$Ge \cite{Lettmann2017}.
In the neutron-rich region, the $B(E2)$ behavior has a smooth decrease toward $N=50$ \cite{Iwasaki16}.
Both the shape transition from spherical to weakly deformed and the coexistence
of different types of deformation might occur in these isotopes \cite{Mukhopadhyay17}.
A rich variety of shapes and shape coexistence in Ge isotopes
provide a challenging testing ground for theoretical models. The Skyrme-Hartree-Fock (SHF)
and Gogny Hartree-Fock-Bogoliubov (HFB)
models imply that most Ge isotopes show the features of soft triaxial deformation \cite{Guo18}.
The self-consistent total-Routhian-surface calculations show there exists the shape phase
transitions from oblate deformation, through triaxial deformation, to prolate deformation
in even-mass $^{64-80}$Ge isotopes \cite{Sheng14,Sarriguren19}. The nuclear density functional
theory investigated the structural evolution from weakly triaxial deformation in $^{74}$Ge
to $\gamma$ soft deformation in $^{78,80}$Ge, and finally to spherical shape in $^{82}$Ge \cite{T20}.
The multi-quasiparticle triaxial projected shell model demonstrates that $^{76}$Ge exhibits
a rigid $\gamma$ deformation in its low-lying states, which is a rare nucleus possessing this
kind of nuclear structure. But its neighboring nuclei such as $^{70,72,74,78,80}$Ge isotopes
show the different $\gamma$-soft features \cite{Bhat21}. Moreover for $^{80}$Ge, because of
subtle balance between quadrupole terms and pairing term in the interaction, each term of
interaction governs two coexisting systems respectively: one for the quasipaticle type and
the other for the collective triaxial type \cite{Verney22}, these interactions determine the
features of $^{80}$Ge.\

In Ref.\cite{Hsieh24}, the authors discussed the general properites of low-lying states of
the even-even Ge isotopes through the interacting boson model (IBM-1). One does not distinct
neutron pairs and proton pairs in IBM-1 \cite{Iachello23}. The calculation results reproduced
the available experimental data, and suggested that there exist the shape transitions from
the mixture of U(5), SU(3) and O(6) symmetry to the mixture of U(5) and O(6) and finally to U(5)
symmetry along the even-even isotopes of $^{64-78}$Ge \cite{Hsieh24}. Meanwhile, the authors
of Ref.~\cite{Duval25} reproduced satisfactorily
the available experimental information on the energy spectum, $E2$ transition and quadrupole
moments for the even-mass $^{68-76}$Ge through the proton-neutron interacting
boson model (IBM-2). In IBM-2, proton bosons and neutron bosons are independently cheated
as different degree of freedom and introduces the mixing of their configurations \cite{Iachello23}.
Furthermore, the energy levels, $E2$ and $M1$ transition properties of even-even isotopes $^{64-68}$Ge
were analyzed through the IBM model with isospin (IBM-3) \cite{Elliott27}.\

Very recently, the shape coexistence and shape transitions in the even-even nuclei $^{66-94}$Ge
were calculated by using the IBM-1 \cite{Nomura2017}, where the authors applied a self-consistent
mean-field method on the basis of the Gogny-D1M energy density functional theory. This calculation
agreed with the known experimental data of these nuclei. However, their calculated energy levels
of the states $E(0^+_2)$ and $E(2^+_2)$ are a little higher than the experimental data especially
for $^{80}$Ge, the reason is that the proton-neutron pairing effects could not be neglected in this
case. On the other hand, the IBM-2 without introducing the configuration mixing has been used to
investigate shape coexistence in some nuclei in the $A\sim100$ mass region \cite{Zhang28,Zhang29},
and in the neutron-deficient isotopes $^{74,76}$Kr \cite{Zhang30}.
The numerical calculations are in good agreement with the recent experimental values for the low-lying
energy spectrum, and the key sensitive quantities such as the quadrupole shape invariants and the $B(E2)$
transition strength branch ratios. In particular, the calculation reproduces the low-lying 0$^+_2$ state
quite well, which is intimately related with the shape-coexistence phenomenon. However, there is short
of a detailed investigation on the nuclear shape and shape coexistence in the exotic nucleus $^{80}$Ge
by IBM-2. In this study, we will discuss the properties of the low-lying states of $^{80}$Ge, especially
the relevant shape coexistence in the framework of IBM-2. Based on the fact that the relative energy
of $d$ neutron boson is different from proton boson's energy, we calculate the energy levels of low-lying states,
and the $B(E2)$ and $\rho^2(E0)$ transition strengths. We also compare the numerical results with the
recent available experimental data. Then, we will describe the shape coexistence phenomena in $^{80}$Ge with IBM-2.\

The structure of this paper is listed as follows. In Sec.~\textrm{\ref{s-2}}, we briefly describe the Hamiltonian,
$E2$ and $E0$ operators used in this study, and also present the criteria adopted for determining the IBM-2 model
parameters. In Sec.~\textbf{\ref{s3}}, we compare the numerical results and experimental data and discuss the
electromagnetic transition properties. Finally in Sec.~\textrm{\ref{s4}}, we give our summary and some remarks.\

\section{Theoretical framework}
\label{s-2}
In IBM-2, the total bosons include proton bosons and neutron bosons, namely satisfy $N=N_\pi+N_\nu$. The
boson creation operators $s^+_{\rho,0}$ and $d^+_{\rho,\mu}$ and the corresponding annihilation operators
$s_{\rho,0}$ and $d_{\rho,\mu}$ constructed the generators of the group U$_\pi\otimes$U$_\nu$ , where $\rho$
represents $\pi$ or $\nu$ and $\mu=-2,...,2$. The product $[N_\nu]\times[N_\pi]$ of symmetric representations
of $U_\pi(6)$ and  $U_\nu(6)$ constitutes the IBM-2 model space.
The IBM-2 Hamiltonian used in this paper has the standard form \cite{Iachello23}
\begin{equation}
\label{Hamil1-1}
\hat{H} = \varepsilon_{d\pi}\hat{n}_{d\pi} +  \varepsilon_{d\nu}\hat{n}_{d\nu}  +
\kappa_{\pi\nu}\hat{Q}_{\pi} \cdot \hat{Q}_{\nu}+\omega_{\pi\pi}\hat{L}_{\pi}\cdot\hat{L}_{\pi}
+\hat{M}_{\pi\nu} \, ,
\end{equation}
where
$\hat{n}_{d\rho} = d^{\dag}_\rho \cdot\tilde{d_\rho} \, $ stands for $d$-boson number operator for
neutron $(\rho= \nu)$ and proton $(\rho= \pi)$, respectively. $\varepsilon_{d\rho}$ is
the energy of the \emph{d}-bosons relative to the $s$ bosons.
$\hat{Q}_{\rho} = (s^{\dag}_{\rho} \tilde{d}_{\rho} +
d^{\dag}_{\rho} {s}_{\rho})^{(2)} +
\chi_{\rho}(d^{\dag}_{\rho}\tilde{d}_{\rho})^{(2)} \, $
denotes the quadrupole operator.
${\chi_\rho}$ occurred in the quadrupole
operator determines the type of the deformation. The third term represents the quadrupole-quadrupole interaction
between proton-boson and neutron-boson with the strength parameter $\kappa_{\pi\nu}$.
The fourth term of Eq.(\ref{Hamil1-1}) denotes the dipole proton-proton interaction with
strength ${\omega_{\pi\pi}}$, where $\hat{L}_\pi$ is the angular momentum operator, which can be explicitly
expressed as $\hat{L}_\pi=\sqrt{10}[d^{\dag}_\pi \cdot\tilde{d}_\pi]^{(1)}$. The last term denotes the Majorana
interaction, its explicit form is
$\hat{M}_{\pi\nu} = \lambda_2(s^{\dag}_{\pi}d^{\dag}_{\nu} -
s^{\dag}_{\nu} d^{\dag}_{\pi})^{(2)} \cdot
(s_{\pi}\tilde{d}_{\nu}-s_{\nu}\tilde{d}_{\pi})^{(2)}$\
$+\sum_{k=1,3}\lambda_{k}(d^{\dag}_{\pi}
d^{\dag}_{\nu})^{(k)}\cdot(\tilde{d}_{\pi}\tilde{d}_{\nu})
 ^{(k)}\, $,
where the strength of Majorana interaction are embodied by the parameters $\lambda_{k}$ (\emph{k}=1,2,3).\

The Hamiltonian in Eq.(\ref{Hamil1-1}) gives rise to four dynamical symmetries $\textrm{U}_{\pi \nu }(5)$,
$\textrm{SU}_{\pi\nu}(3)$, $\textrm{O}_{\pi\nu}(6)$, and $\textrm{SU}^{\ast}_{\pi \nu}(3)$, which correspond
to a spherical, an axially symmetric, a $\gamma$-unstable, and a triaxial deformed shape respectively.
For certain values of the model parameters, the Eq.(\ref{Hamil1-1}) can reduce to contain
only one kind of dynamical symmetry \cite{Cejner32}. The $B(E2)$ transition strengths and the $\rho^2(E0)$
values between 0$^+$ states could be used to search the signatures of shape coexistence. In IBM-2, the \emph{E}2
transition matrix element is defined as follows
\begin{equation} \label{B-Operator}
B(E2,J{\rightarrow}J') =
\frac{1}{2J+1}|{\langle}J'\|\hat{T}^{(E2)}\|J\rangle|^2\, ,
\end{equation}
where the $E$2 transition operator $\hat{T}^{(E2)}$
is given through the quadrupole operator $Q_\rho$ as $\hat{T}^{(E2)}=e_{\pi} \hat{Q}_{\pi} + e_{\nu}
\hat{Q}_{\nu}$. $J$ and $J'$ are the initial and final angular momenta, respectively.
$e_\nu(e_\pi)$ represents the effective charge of
neutron (proton) bosons, one can determine the effective charges by fitting the experimental data.\

The $E0$ transition matrix element $\rho$ in the IBM-2 is defined as
\begin{equation} \label{p0-Operator}
\rho(E0,J{\rightarrow}J') =\frac{Z}{e R^2}[
\beta_{0\pi}{\langle}J'\|\hat{T}^{(E0)}_\pi\|J\rangle+
\beta_{0\nu}{\langle}J'\|\hat{T}^{(E0)}_\nu\|J\rangle]\ ,
\end{equation}
where $R=1.2A^{1/3}$fm, and $\beta_{0\pi(\nu)}$ is the
so-called proton (neutron) monopole boson effective charge in unit of $e$fm$^2$.
The $E$0 transition operator is written as
$\hat{T}^{(E0)}=\beta_{0\pi}\hat{T}^{(E0)}_\pi + \beta_{0\nu}\hat{T}^{(E0)}_\nu
=\beta_{0\pi} \hat{n}_{d\pi} + \beta_{0\nu} \hat{n}_{d\nu}$,
where the $\hat{n}_{d\rho}$ is the same in Eq.(\ref{Hamil1-1}).\

$^{80}$Ge is composed of $N=48$ neutrons and $Z=32$ protons, and locates at $Z=28$, $N=50$ major shell.
We take the doubly magic nucleus $^{78}$Ni at $Z=28$ and $N=50$ as an inert core for the description of $^{80}$Ge.
In this case, there are two proton bosons outside the $Z = 28$ shell which are particle-like,
while one neutron bosons outside the $N = 50$ shell in $^{80}$Ge which is hole-like.
The microscopic picture demonstrates that the valence neutrons and protons occupy different orbitals
when they are added to $^{78}_{28}$Ni$_{50}$ core \cite{Nowacki04,G33}. The four valence protons are distributed
among the $fp$ orbitals. The two hole-like valence neutrons occupy the $g_{9/2}$ orbital \cite{Verney22,G33,Rotter1990}.
In order to consist with the microscopic description and remove some of the degeneracies,
we use the different energies $\varepsilon_{d\pi}\neq\varepsilon_{d\nu}$ for $d$ proton and neutron
boson in the same way as in Refs. \cite{Zhang28,Dejbakhsh34}. In general, the parameters $\varepsilon_{d\rho}$
and $\kappa_{\pi\nu}$ are mainly used to reproduce the energy levels of low-lying states with positive parity.
The values of $\varepsilon_{d\rho}$ mostly contribute to the spectrum of $\textrm{U}(5)$ nuclei. However
$\kappa_{\pi\nu}$ mainly characterizes the properties of deformed
nuclei. The structure parameters $\chi_\pi$ and $\chi_\nu$
occurred in quadrupole operators are used to describe the $B(E2)$ transition properties. Only the dipole
interaction term $\hat{L}_{\pi}\cdot\hat{L}_{\pi}$ is explicitly
considered in the Hamiltonian because of only one hole-like neutron
boson outside the $N=50$ shell in $^{80}$Ge. $\hat{L}_{\pi}\cdot\hat{L}_{\pi}$ plays an important role on
the description of rotational energy levels \cite{Nomura35,Nomura36,Zhang2017}. The parameter $\omega_{\pi\pi}$
can be used to tune the order of the 2$^+_2$ state and 4$^+_1$ state. The Majoranan parameters mainly
influence the mixed symmetry states, in order to reduce the number of the free parameters in Hamiltonian,
for simplicity we take $\lambda_2=0$ and $\lambda_1=\lambda_3$ in this study.\

The IBM-2 parameters are determined to reproduce the the experimental data for $^{80}$Ge: $\varepsilon_{d\pi}=0.315$MeV,
$\varepsilon_{d\nu}=1.080$ MeV, $\kappa=-0.150$ MeV, $\chi_{\pi}=-1.200$, $\chi_{\nu}=0.900$,
${\omega_{\pi\pi}}=0.063$ MeV, and $\lambda_1=\lambda_3=0.800$ MeV in $^{80}$Ge. We numerically diagonalized
the IBM-2 Hamiltonian by the NPBOS code \cite {Otsuka37}. The obtained IBM wave functions are our starting point
and can be used to compute the electromagnetic properties.\

\section{Results and discussion}
\label{s3}
\begin{figure}
\begin{center}
\includegraphics[width=8cm]{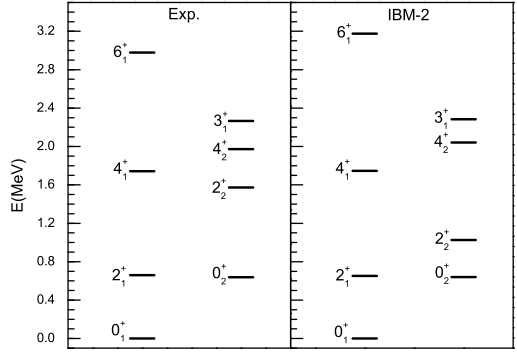}
\caption{
\label{fig1}
The energy scheme for low-lying states of $^{80}$Ge with positive-parity , the left panel shows the
experimental data and the right panel denotes the calculated results from IBM-2. The experimental
energy levels are taken from Ref.\cite{Gottardo11,Verney22}.
\label{figl}}
\end{center}
\end{figure}

The calculated results of the low-lying energy levels compared with the corresponding available experimental
data are shown in Fig.~\ref{fig1}. The experimental values are taken from Refs.\cite{Gottardo11,Verney22}.
Each panel includes two different parts: the yrast band up to the 6$^+$ state and the nonyrast,
low-spin, positive-parity levels. Fig.~\ref{fig1} shows that the calculated energy levels from IBM-2 for the low-lying
states agree with experimental data very well. The experimental energy levels of the yrast states are
reproduced precisely by the theoretical calculations.
At the same time, the calculated ordering of the nonyrast states consists with the experimental data, although
the theoretical prediction of the 2$^+_2$ state is lower than the experimental value.
In particular, the calculated result of the first excited 0$^+_2$ state is almost equal to the experimental
data of 639(1) keV, which is lower than the 2$^+_1$ state.

The energy ratio $R_{4/2}=E(4^+_1)/E(2^+_1)$ of $2^+_1$ state and $4^+_1$ is a well-known observable
to measure the extent of quadrupole deformation. $R_{4/2}$ reaches the limit of 2.00 for
the $\textrm{U}_{\pi \nu }(5)$ dynamical symmetry (the spherical vibration), 2.50 for the $\textrm{O}_{\pi\nu}(6)$
dynamical symmetry (the  $\gamma$-unstable rotor), and the maximum 3.33 for the $\textrm{SU}_{\pi \nu }(3)$
dynamical symmetry (the axially rotor) \cite{Cejner32}. The experimental result of $R_{4/2}$ is 2.64 for $^{80}$Ge,
and the calculated value is 2.68. Both the experimental and theoretical value of R$_{4/2}$ predict that $^{80}$Ge
has a mostly typical $\gamma$-soft triaxial feature. At the same time, from Fig.~\ref{fig1} one can see that both
of the experimental and calculated energy levels of the 2$^+_2$ state lie below the corresponding
4$^+_1$ state, and they form a pair of 2$^+_2$ and 4$^+_1$, which also indicates $^{80}$Ge
exhibits a characteristic of O$_{\pi\nu}(6)$ symmetry because the fact of the second 2$^+$
state below the 4$^+_1$ state is the manifestation of a $\gamma$-soft spectrum.
However, the appearance of the 4$^+_1$ state at an energy of nearly 2.5 times that of 2$^+_1$ level
alone does not uniquely determine the O$_{\pi\nu}$(6) structure \cite{Casten38}.
In the O$_{\pi\nu}$(6) limit of IBM-2, the 2$^+_2$ state and 4$^+_1$ state belong to the $\tau=2$ multiplet, but
the 6$^+_1$, 0$^+_2$, 3$^+_1$, 4$^+_2$ states belong to $\tau=3$ multiplet. As a consequence, the 0$^+_2$ state
($\tau=3$ multiplet) locates at much higher energy level and can decay to the second 2$^+$ state
with $\tau=2$ rather than to the 2$^+_1$ state. But the 0$^+_2$ state of $^{80}$Ge actually
lies at lower energy level than the 4$^+_1$, 2$^+_2$ states, even below the 2$^+_1$
level both in experiment and theory. From the above discussion it is clear that the $^{80}$Ge
is not a typical $\gamma$-soft nucleus, at least deviates from the pure O$_{\pi\nu}$(6) limit,
although the yrast states have approximately the $\gamma$-soft rotor picture. More importantly, it is an
important evidence of shape coexistence if a deformation state occurs near the almost spherical ground
state or much lower than the first-excited 2$^+$ state \cite{Andreyev05}. Therefore, both the experimental
and theoretical energy levels imply that shape coexistence occurs in $^{80}$Ge.\

$B(E2)$ transition probability and its branching ratios can also
give important information on the nuclear structure. Unfortunately, only absolute $B(E2)$ transition
strengths of 2$^+_1 \rightarrow 0^+_1$, and 2$^+_2 \rightarrow 0^+_1$ in $^{80}$Ge
have been observed so far. However, one can further explore shape coexistence in $^{80}$Ge based on
the other key sensitive quantities \cite{Casten38,Zhang39}. To calculate the $E2$ transition strengths,
the effective charges of proton and neutron bosons were determined to reproduce the experimental data
of $B(E2, 2^+_1\rightarrow0^+_1)$ and $B(E2, 2^+_2\rightarrow0^+_1)$.
By fitting the experimental data of the $B(E2,2^+_1\rightarrow0^+_1)=200(26)$ $e^2$fm$^4$, we obtain
the $e_\nu=13.9$, and $e_\pi=6$ $e$fm$^2$ for $^{80}$Ge. The effective charge of neutron boson is much
larger than proton boson's probably due to the effect of a valid proton midshell around $Z=34$.
For the protons, the state space beyond the $Z=20$ shell closure and up to $Z=32-34$ is indeed made
of the full $pf$ shell \cite{Nowacki04}, which might lead to a very valid proton subshell closure at $Z=32$
and $34$. The other reason is that the parameters $e_\nu$ and $e_\pi$ incorporate a (length)$^2$ factor,
simultaneously, the neutrons are occupying higher shells than protons in $^{80}$Ge \cite{Hamilton40}.\

\begin{table}[ht]
\begin{center}
\caption{ \label{tab1} The experimental and calculated $B(E2)$ values
(in $e^2$fm$^4$) and $\rho^2(E0)$ values in $^{80}$Ge are listed. We take the experimental data from Refs.
\cite{Gottardo11,Verney22}.}
\footnotesize
\begin{tabular*}{90mm}{cccccccccc@{\extracolsep{\fill}}cccccccccccccc}
\toprule &$B(E2, 2^+_1\rightarrow0^+_1)$  & $B(E2, 2^+_2\rightarrow0^+_1)$ & $\rho^2(E0, 0^+_2\rightarrow0^+_1)$  \\
\hline
Exp.    & 200(26) & 23(7) &  \\
Cal.     & 200.0   & 21.3  & 0.001  \\
\hline
\hline
\end{tabular*}
\end{center}
\end{table}

The calculated $B(E2)$ transition strengths comparing with the recent experimental values are listed
in Table \ref{tab1}. The theoretical calculations are in consistence with the experimental data quite nicely.
The calculated transition strength of $B(E2, 2^+_2\rightarrow0^+_1)$
is in agreement with the experimental value within the experimental uncertainty. In the IBM, the key sensitive
quantities $R_1=B(E2, 2^+_2\rightarrow2^+_1)/B(E2, 2^+_1\rightarrow0^+_1)$ and
$R_2=B(E2, 2^+_2\rightarrow0^+_1)/B(E2, 2^+_2\rightarrow2^+_1)$  are usually considered as one of
the most crucial available structure indicators \cite{Casten38} to distinguish the dynamical symmetry
limits. The U(5) symmetry is realized when $R_1= 1.40$ and $R_2= 0.011$, while it is the O(6) symmetry
when $R_1= 0.79$ and $R_2= 0.07$ \cite{Stachel1982}. The calculation result of $B(E2, 2^+_2\rightarrow2^+_1)$
is 187.13 $e^2$fm$^4$. The calculated $R_1$ and $R_2$ are 0.94 and 0.11 respectively, which are much
closer to O(6) symmetry. Obviously, the predict ratios of $R_1$ and $R_2$ are consistent with the
character of the yrast states, but do not match with the feature of the nonyrast states.
Thus, the above result has confirmed the existence of shape coexistence in $^{80}$Ge.\

One can obtain valuable information from the electric monopole transition strengths $\rho^2(E0)$ on the
excited 0$^+$ states of different features coexisting in the same nucleus \cite{Mantica40,Bouchez41,Brown2017}.
In order to further understand the properites of shape coexistence in $^{80}$Ge, we
calculate $\rho^2(E0, 0^+_2\rightarrow0^+_1)$ transition strength. Since the experimental
data about $E0$ transition is still scarce in $^{80}$Ge, we choose the parameters
$\beta_{0\nu}$ and $\beta_{0\pi}$ as the values derived in ref.\cite{Barrett40} from a
detailed analysis of $E0$ transition in O(6)-like nuclei, namely, $\beta_{0\nu}=0$
and $\beta_{0\pi}=0.20$ $e$fm$^2$. The calculated transition strength is also listed
in Table \ref{tab1}. Because the $E0$ operator is proportional to $\hat{n}_d$,
no $E0$ transitions occur in the U(5) dynamical limit \cite{Leviatan43}. Within the O(6) limit,
the selection rules are $\triangle\sigma=0,\pm2$, $\triangle\tau=0$, so the $0^+_2\rightarrow0^+_1$ transition
is forbidden \cite{Wood45}. The present calculation value of $\rho^2(E0)$ is
comparable with those observed in $^{72}$Ge, $^{102}$Pd and $^{120}$Xe \cite{Mantica40,Wood45},
which implies that different nuclear shapes coexist in $^{80}$Ge.\

On the other hand, the choice of the parameters to reproduce the properties of the low-lying states might give us
a clue to understand shape coexistence in nuclei. Recalling the best fit parameters in the present calculation,
we found that the $\varepsilon_{d\rho}$ is much larger than $\kappa_{\pi\nu}$,
which reflects that $^{80}$Ge mainly exhibits the character of spherical vibration or U(5) dynamical symmetry.
At the same time, the structure parameter of the quadrupole operator $\chi_\pi=-1.200$, and $\chi_\nu=0.900$ were
adopted in this paper. The sum $\chi_\pi+\chi_\nu=-0.3$ indicates that $^{80}$Ge nucleus is close to the O(6)
dynamical symmetry or $\gamma$-soft in IBM. As mentioned above, combing the information from the best fit parameters
and the properties of the low-lying states, the physical picture from the IBM point of view is clear: both collective
spherical and $\gamma$-soft vibration structures coexist in $^{80}$Ge. Microscopically, the recent shell model
calculations in the $pfgd$ model space suggest that the tensor force play an important role on setting up a shape
coexistence environment and the tensor effect changes dynamically with orbital occupation and spin \cite{Kaneko2017}.
For $^{80}$Ge, many neutrons occupying $g_{9/2}$ orbital reduce the proton $f_{7/2}-f_{5/2}$ gap, much more particle-hole
excitations occur over the gap, which lead to much stronger shell evolution \cite{Tsunoda2014}. The other studies have
clearly shown that the $\nu s_{1/2}$ shell drops in energy and becomes almost degenerate with the lower-lying $\nu d_{5/2}$
shell at $Z=32$ \cite{Gottardo11}. Therefore, neutron pair excitations across $N=50$ are likely to include both orbitals
which result in significant configuration mixing. The deformation and change of shell structure driven by the
combination of the tensor force and changes of major configurations can occur and can enhance shape coexistence
in $^{80}$Ge.\

\section{Conclusion}\
\label{s4}
In summary, we discussed the properties of the low-lying states, especially the relevant
shape coexistence in $^{80}$Ge near one of most neutron-rich doubly magic nuclei at $N=50$
and $Z=28$. Based on the different relative energy for $d$ proton boson neutron boson, $i.e.,$
$\varepsilon_{d\pi}\neq\varepsilon_{d\nu}$, the low-lying positive parity states consist
with experimental data very well in IBM-2. More importantly, the calculated energy level of
the first excited 0$^+_2$ state, which associates with the shape coexistence phenomenon, is
almost equal to the experimental data at 659 keV, which is lower than the 2$^+_1$ state.
Both the experimental and theoretical energy spectrum indicated the shape coexistence exists
in $^{80}$Ge structure, although the value of the characteristic ratio of $R_{4/2}$ suggests
that $^{80}$Ge is a mostly typical $\gamma$-soft triaxial feature.\

The calculated $B(E2)$ transition strengths agreed with the experimental data within the experimental
uncertainty. The key sensitive quantities do not match with the feature of the nonyrast states,
which demonstrates the different property of $^{80}$Ge compared with its energy spectrum structure.
Therefore, the above result has just confirmed the existence of shape coexistence in $^{80}$Ge.
Furthermore, the $\rho^2(E0, 0^+_2\rightarrow0^+_1)$ transition strength has been calculated. The
theoretical result of $\rho^2(E0, 0^+_2\rightarrow0^+_1)$ transition also indicates the different
nuclear shapes exist in the same time in $^{80}$Ge.\

The best fit values of $\varepsilon_{d\rho}$ is much larger than $\kappa_{\pi\nu}$, which implies that
$^{80}$Ge has the property of U(5) dynamical symmetry. While the sum $\chi_\pi+\chi_\nu=-0.3$ indicates
that $^{80}$Ge is close to the $\gamma$-soft or O(6) dynamical symmetry in IBM. Combing the results of
the best fit parameters in present calculations and the properties of the low-lying states, we found that
both collective spherical and $\gamma$-soft vibration structures coexist in $^{80}$Ge from the IBM point
of view. However, the experimental information on $E2$ and $E0$ transition from 0$^+_2$ state to other
states in $^{80}$Ge is still scarce. As a result, our theoretical analysis for the associated 0$^+_2$
level might be incomplete. More theoretical calculations and experimental investigations on these aspects are needed.\

We thank  professors Y. X. Liu, G. L. Long and C. W. Shen for helpful
discussions.  This work is supported by the National Natural Science Foundation of China
under grant numbers 11475062, 11647306 and 11147148.

\clearpage
\end{document}